\documentclass[USenglish,twocolumn]{article}

\usepackage[utf8]{inputenc}				
\usepackage[big,online]{dgruyter}	
\usepackage{lmodern} 
\usepackage{microtype}
\usepackage[numbers,square,sort&compress]{natbib}

\theoremstyle{dgthm}

\theoremstyle{dgdef}

\usepackage{siunitx}

\begin{document}

	\articletype{Research Article}
	\received{September 12, 2025}
  \startpage{1}

\title{Impact of aberrations in SLM-based far-field holography}

\runningtitle{Aberrations in far-field holograms}

\author*[1]{Markus Zimmermann}
\author[1]{Andreas Brenner}
\author[1]{Tobias Haist} 
\author[1]{Stephan Reichelt} 
\runningauthor{M.~Zimmermann et al.}
\affil[1]{\protect\raggedright 
University of Stuttgart, Institute of Applied Optics, Stuttgart, Germany, e-mail: zimmermann@ito.uni-stuttgart.de}
	

\abstract{We use camera-in-the-loop calibration to calibrate a phase-only spatial light modulator (SLM) in a far-field hologram setup. The recorded intensity distributions achieve a high degree of consistency with the calculated results, indicating a precise calibration and sufficient modeling of the most prominent aberrations. In this work, we discuss the modeled aberrations and examine the improvement or loss in image quality and diffraction efficiency that is obtained by including or excluding the modeled aberrations in the calibration. We further show the influence of aberrations on speckle-reduced holograms and evaluate the speckle contrast. }


\maketitle

\section{Introduction} 

Phase-only spatial light modulators (SLM) feature a wide range of applications, as they offer the ability to dynamically alter the phase of an incoming wavefront. SLMs are used in biology experiments \cite{yangReviewLiquidCrystal2023} or quantum technology as dynamic optical traps \cite{schroffAccurateHolographicLight2023}, laser material processing \cite{10.1117/12.3023102}, and as display for holographic display concepts \cite{blancheHolographyFuture3D2021, javidiRoadmapDigitalHolography2021, piReviewComputergeneratedHologram2022}. An extensive overview of applications and basic calibration techniques can be found in the manufacturer's application notes \cite{holoeyephotonicsagHOLOEYESpatialLight2023, RecentResearchUsing, hamamatsuphotonicsPublishedPapersHamamatsu}.
As phase-only computer generated holograms (CGHs) have a higher light efficiency than amplitude CGHs, phase-only SLMs are preferred for a lot of applications over amplitude SLMs. Phase-only CGH calculation is often performed using some kind of iterative procedure \cite{gerchbergPracticalAlgorithmDetermination1972, pasienskiHighaccuracyAlgorithmDesigning2008}. In recent years gradient descent schemes have been proposed and have gained popularity for the CGH calculation \cite{chakravarthulaWirtingerHolographyNeareye2019}. However, apart from different calculation methods, the achieved reconstruction quality of the hologram depends heavily on the calibration of the optical system.
Camera-in-the-loop (CITL) calibration, originally introduced for the calibration of SLMs in near-eye displays, offers a simple method to achieve calibration precision that exceeds the classic calibration methods \cite{pengNeuralHolographyCameraintheloop2020, chakravarthulaLearnedHardwareintheloopPhase2020}. 
In this work, the influence of the different boundary conditions, such as the spatially varying illumination, the pixel cross talk (also known as the fringing field effect), the non-linear phase modulation and the phase aberrations, are investigated.

\section{Methods}

\begin{figure}
\begin{center}

\includegraphics[width=83mm]{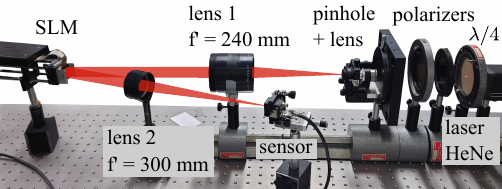}

\end{center}
\caption 
{ \label{fig:aufbau}
Optical setup with the sketched optical path. The SLM is illuminated with collimated light and the sensor is placed in the back focal plane of lens 2. Lens 1 has an internal aperture, which limits the beam diameter to a size which is slightly larger than the SLM. The polarizers and the quarter wave plate ensure the correct polarization for the SLM and allow for adjusting the intensity of the illumination.} 
\end{figure}

\subsection{Camera-in-the-loop (CITL) calibration}
Our CITL calibration uses a neural network to create a digital twin of the physical setup \cite{shimobabaDeepLearningComputationalHolography2022}. This neural network is used in a gradient descent scheme to calculate phase-only holograms, which are displayed on the SLM and the intensity pattern created by the hologram is recorded by a camera. The camera recordings are then used to train the neural network to adjust the weights in order to minimize the difference between the predicted and recorded camera image.

Although neural networks are typically seen as black-boxes where the weights of the different layers cannot be linked to any specific behavior, digital twins for SLM calibration can be created as a physically interpretable neural network using Fourier optics, where each layer corresponds to a specific optical aberration \cite{zimmermannImprovedFarField2025}. Often, physically interpretable implementations are expanded with a small black-box model, to account for aberrations that are not considered by the implemented aberrations \cite{pengNeuralHolographyCameraintheloop2020, choiTimemultiplexedNeuralHolography2022}.

\subsection{Modeled aberrations}
We investigate a far-field hologram setup as is shown in \autoref{fig:aufbau}. The intensity in the far-field can be obtained from the phase-only hologram by converting the input phase at the SLM to a complex wavefield using a uniform approximation of the illumination amplitude and propagating the wavefield into the far-field by calculating the fast Fourier transform. After the fast Fourier transform the intensity is computed by taking the square of the absolute from the wave field. Using this naive assumption about the boundary constraints, all aberrations that can occur in the system are neglected and a system with a uniform illumination amplitude and perfectly flat illumination wavefront and no aberrations is assumed.

\begin{figure*}
\begin{center}
\includegraphics[width=16.5cm]{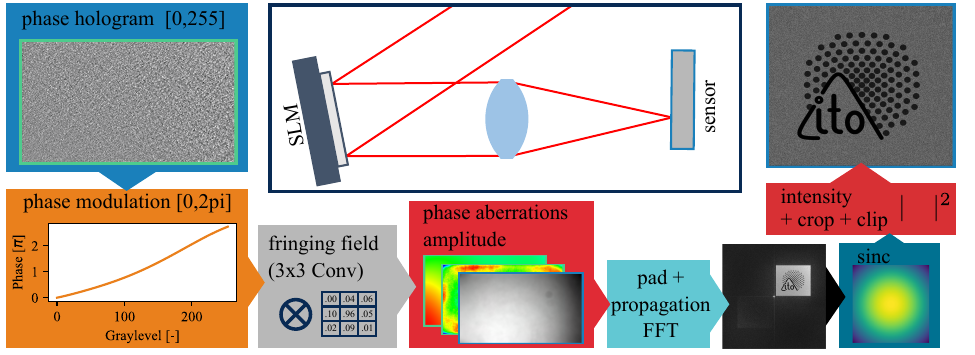}
\end{center}
\caption 
{ \label{fig:model}
Visualization of the structure of the digital twin and the schematic of the optical setup in the middle. The values from the phase-only hologram are converted to the actual phase delay introduced by the SLM. Next, these values are altered by the fringing field effect. Phase aberrations are added and the phase is converted to a complex wave field by adding the illumination amplitude. The wave field is zero-padded to maintain a uniform pixel size after propagation using a fast Fourier transform. The far field is attenuated by a sinc distribution and finally the intensity is calculated and the target area is cropped from the far field.} 
\end{figure*}

We implemented the most prominent aberrations that can affect the quality of the intensities generated by a far-field hologram. The flow of information through the modeled system can be seen in \autoref{fig:model}. We did not include the quantization of the phase values, since there is only minor quality loss for a quantization to 8-bit \cite{heOptimalQuantizationAmplitude2021, morenoDiffractionEfficiencyStepped2020}.

Here we give an overview of the incorporated aberrations and how they can be modeled. In the results, we examine the impact on image quality, diffraction efficiency, and speckle contrast for speckle-reduced far-field intensities when these aberrations are either included or omitted.

\subsubsection{Phase modulation}
In order to display a phase-only hologram and achieve a high reconstruction quality, the relation between the phase modulation and the displayed gray value must be known. Manufactures usually calibrate the SLM and provide an internal look-up table to achieve a linear relation between phase modulation and displayed gray value. This look-up table can be adjusted for different wavelengths by the end user. 
However, the relationship between phase modulation and the gray value displayed will also change depending on the illumination angle and can depend on the temperature or deformations from mounting forces and can therefore also be spatially non-uniform \cite{reicheltSpatiallyResolvedPhaseresponse2013}.
We model the relationship between phase modulation and the gray value as a global relation by using a 10th-order polynomial:
\begin{equation}\label{eq:p2v}
\varphi_{out}(u_{in}) = \sum_{i=1}^{10}{a_i\cdot(u_{in}-b_i)^{i}}.
\end{equation}

\subsubsection{Fringing field effect}
The fringing field effect describes the crosstalk between neighboring pixels. For liquid-crystal-based SLMs, the liquid crystals are oriented using electric fields. As the electric fields are not perfectly confined to one pixel, the orientation of the liquid crystals can be affected by the electric field from the neighboring pixel. This way the resulting phase delay can depend on the value from the neighboring pixels. Crosstalk becomes more severe as the pixels get smaller, since the relative overlap of the electric fields increases. In SLMs, smaller pixel sizes are preferred, as they lead to increased diffraction angles. The fringing field effect is stronger for binary patterns, where a high gradient of pixel values exists. The fringing field effect has direct influence on the diffraction efficiency \cite{morenoSimpleMethodEvaluate2021}.
We use a convolution with a 3x3 kernel, to model the fringing field effect \cite{lingelOptimizingDiffractionEfficiency2013, moserModelbasedCompensationPixel2019}.

\subsubsection{Phase aberrations}
Static phase aberrations can occur due to form deviations in the SLM surface, either the mirrored backplane or the glass cover. Apart from the SLM, they can also arise from non-ideal lenses, which introduce phase aberrations. We model phase aberrations with low spatial frequencies by using Zernike polynomials and phase aberrations with high spatial frequencies by adding a phase distribution, where every pixel can have a different value.

\subsubsection{Illumination amplitude}
The illumination can be derived through external measurements, like taking an image of the SLM surface in the setup. However, every external measured amplitude distribution needs to be aligned with the corresponding SLM pixel. Some indirect measurement methods have been proposed, such as reconstructing the illumination amplitude from measurements of the focal plane \cite{vedrenneLaserBeamComplex2014}. In an ideal scenario, the illumination amplitude could be modeled by a Gaussian distribution because the light originates from a pinhole and the collimated beam is much larger than the SLM. This neglects higher spatial frequency deviations in the illumination, which can occur from dirt on the lenses or the SLM surface, or undesired reflections in the system. Therefore, we model the illumination amplitude similar to the high-frequency phase aberrations by using a distribution that can have a different value for each pixel.

\subsubsection{Fill factor and rectangular pixels}
Due to the rectangular shape of the pixels of the SLM, the resulting far-field intensity is attenuated with the Fourier transform of the rectangular pixel shape, which corresponds to a sinc distribution. If the fill factor were \SI{100}{\%}, the width of the sinc would correspond to the first diffraction order, but since the fill factor is lower, the width is larger and the attenuation is smaller. This effect is less prominent in areas close to the optical axis \cite{haistHolographyUsingPixelated2015a}.

\subsection{Optical setup and hardware}
The optical setup for the CITL calibration is a simple far-field holographic setup as depicted in \autoref{fig:aufbau}. The SLM is illuminated with a collimated laser beam (HeNe Laser Thorlabs HNLS008L-EC). The SLM is slightly tilted. In this way a beam splitter is omitted. In the back focal plane of lens 2, the sensor is positioned to record the far field of the SLM.
The SLM used is a HoloEye Pluto with 1920 x 1080 pixels, a pixel pitch of \SI{8}{\micro\metre} and a fill factor of 87\%. The image sensor is a monochrome Ximea XiC with the Sony IMX250 chip with 2464 x 2056 pixels and a pitch of \SI{3.45}{\micro\metre}. The sensor is positioned such that the zeroth diffraction order falls onto one corner of the sensor. The target area used for CITL calibration excludes the zeroth diffraction order and has an area of 2300 x 2000 pixels on the image sensor. The model is trained during the CITL procedure using the mean squared error (MSE) as the loss metric.

\subsection{Measurement of diffraction efficiency and speckle contrast}
We determine the diffraction efficiency by measuring the intensity in the first diffraction order from an optimized blazed grating and the intensity in the zeroth diffraction order when nothing is displayed on the SLM, i.e.
\begin{equation}\label{eq:diff_rel}
\eta = \frac{I_{1,\,\mathrm{Grating}}}{I_{0,\,\mathrm{without}}}.
\end{equation}
To measure diffraction efficiency, we cover the SLM border with a mask and only measure light from the active SLM area.

The speckle contrast is defined as the standard deviation of a uniform illuminated area divided by its mean value \cite{goodmanSpecklePhenomenaOptics2020} 
\begin{equation}\label{eq:speckle_contrast}
C = \frac{\sigma_I}{\bar{I}}.
\end{equation}
We used an area of 900 by 900 pixels in the center of the target area together with the speckle suppression method of Zimmermann et al. \cite{zimmermannImprovedFarField2025}.

\section{Results}

\begin{figure}
\begin{center}
\includegraphics[width=83mm]{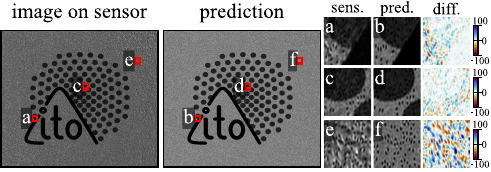}
\end{center}
\caption 
{ \label{fig:compare}
Comparison of the image on the sensor with the prediction as it is calculated from the calibration with the full model. The close ups show a low deviation between the recording and the prediction with a good match of the speckle positions. The difference between prediction and recording are the largest in the upper right corner (e and f), which has the largest distance to the optical axis.} 
\end{figure} 

\begin{figure*}
\begin{center}
\includegraphics[width=165mm]{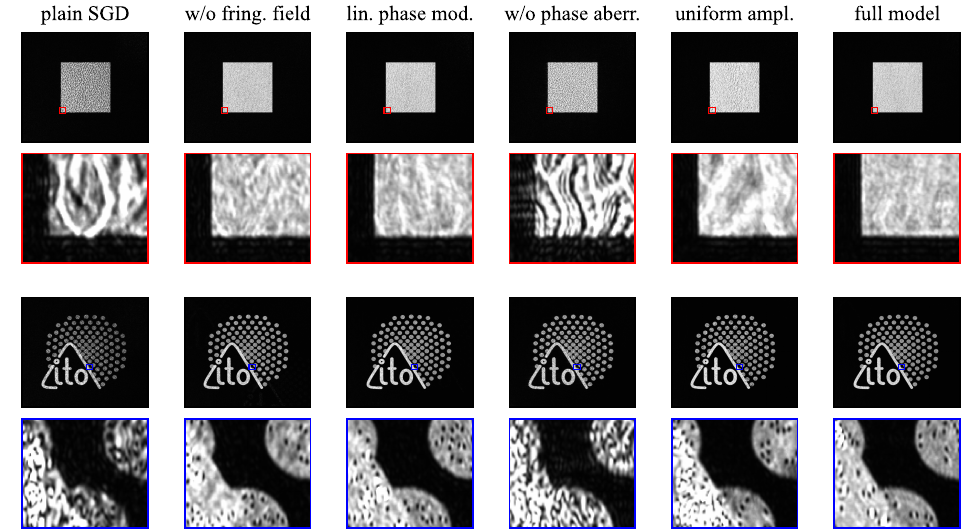}
\end{center}
\caption 
{ \label{fig:results}
Comparison of the recorded intensities in the far field for the different calibration models. The upper row contains a motif where speckle suppression was applied during calculation while the lower row displays a motif without speckle suppression.} 
\end{figure*}

The metrics for the different calibrations with one type of aberration excluded from the calibration model are given in \autoref{tab:Table1}. Two recorded images are shown in \autoref{fig:results}. The peak signal-to-noise ratio (PSNR) is the best for the full model and the lowest, if no aberrations are included as in the plain stochastic gradient descent (SGD). The highest quality difference stems from the phase aberrations. If these are excluded, the structural similarity index measure (SSIM) is as low as in the plain SGD and the speckle contrast is the second highest after the plain SGD. 

The gain in image quality highly depends on the aberrations that are present in the optical system. If there were no phase aberrations in the system, there should be no differences in the metrics between the full model and the model without phase aberrations. However, because the SLM itself typically has some kind of form deviation in the reflective backplane, optical systems with SLMs contain some degree of phase aberrations.

Interestingly, the diffraction efficiency is the highest for the calibration model without the fringing field effect. This is in contrast to research suggesting that the diffraction efficiency is the highest, if the fringing field effect is correctly determined \cite{lingelOptimizingDiffractionEfficiency2013, morenoSimpleMethodEvaluate2021}. However, in our case, the values for the fringing field effect were determined in order to minimize the difference between a target and the recorded pattern, while in other research, the values for the fringing field effect were determined by directly maximizing the diffraction efficiency. Another reason might be that the SGD used for optimizing the blazed grating does not work well with the model of the fringing field effect and the diffraction efficiency drops during optimization.

\begin{table*} [!ht]
\centering
\caption{Calculated PSNR and SSIM of the recorded intensity against the target intensity, the measured speckle contrast and the measured relative efficiency. Different aberrations were excluded from the digital twin, to investigate the influence of the aberration in regards to image quality and diffraction efficiency.}
\begin{tabular}{ccccccc}
 & \parbox{2cm}{\centering plain SGD}& \parbox{2cm}{\centering w/o fringing field effect}  & \parbox{2.5cm}{\centering linear phase mod.} & \parbox{2cm}{\centering w/o phase aberrations} & \parbox{2cm}{\centering uniform illumination} & \parbox{2cm}{\centering full model}	\\ \midrule
PSNR     & 14.90      & 20.24      & 20.67      & 15.69      & 19.76    & \textbf{21.23} \\
SSIM    & 0.16      & 0.22      & 0.24      & 0.16      & \textbf{0.27}    & 0.26 \\
$C$ &   0.45      & 0.15      & 0.14      & 0.37      & 0.15    & \textbf{0.11} \\
$\eta$      & 0.68      & \textbf{0.72}      & 0.70      & 0.71    & 0.70  & 0.70	\\

\end{tabular}
\label{tab:Table1}
\centering
\end{table*}

\subsection{Fringing field effect}
\begin{figure}
\begin{center}
\includegraphics[width=83mm]{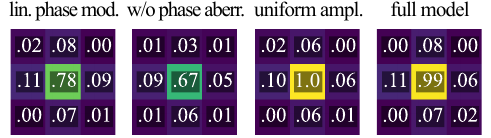}
\end{center}
\caption 
{ \label{fig:fringe}
Comparison of the learned 3x3 kernels to model the fringing field effect.} 
\end{figure} 
The learned kernels for the different calibrations are given in \autoref{fig:fringe}. It can be seen that the diagonal values are the lowest, ranging from 0 to 0.02. This agrees well with the theory that the fringing field effect originates from the electric fields of the neighboring pixels. Since diagonal pixels share only a small edge with the central pixel, the influence of diagonal pixels on the central pixel value is low. The values in the left pixels in the middle columns have the highest value after the central pixels. This might come from the optical setup and the tilt of the SLM. The phase influence might therefore come from the fact that some of the reflected light passes through the neighboring pixel after it was reflected on the backplane, and the phase delay depends to some degree on the value of that pixel.

While the calibrations with uniform amplitude and with the full model yield a central value of one or close to one, the calibrations with linear phase modification and without phase aberrations yield values of 0.78 and 0.67, respectively. For the calibration with the linear phase modification, this restricts the maximum phase delay to 0.78 $\lambda$, which is in agreement with the other maximum phase delays presented in \autoref{fig:phase_mod}. The calibration without phase aberrations shows the lowest value in the center. However, the results in \autoref{fig:phase_mod} show a higher phase delay for this calibration than for the other calibrations. Together with the phase modification from the calibration without phase aberrations, the reduced center of the kernel results in a total phase delay of 0.83 $\lambda$, which is in the range of the other calibrations.

\subsection{Phase modulation}
\begin{figure}
\begin{center}
\includegraphics[width=83mm]{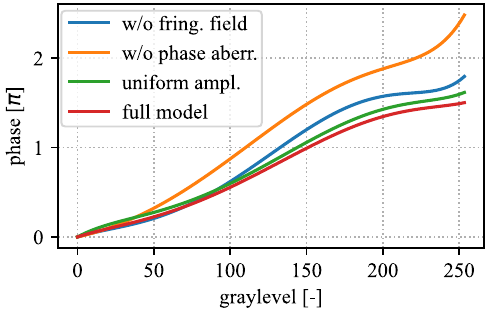}
\end{center}
\caption 
{ \label{fig:phase_mod}
Learned phase modulation in dependency of the greylevel displayed on the SLM. The curve from the calibration without phase aberrations has a value greater than $2\pi$. However, the phase is convolved with the kernel from the fringing field effect and this calibration has learned a lower value in the kernel, therefore the maximum phase delay is less than $2\pi$.} 
\end{figure}

The results of the non-linear phase modulation are shown in \autoref{fig:phase_mod}. A slight non-linearity can be observed and a maximum phase delay between 0.75 and 0.83 $\lambda$, considering the reduction of the phase delay from the calibration without phase aberrations due to the convolution with the kernel of the fringing field effect.
If the maximum phase delay was higher, the diffraction efficiency would increase \cite{morenoModulationLightEfficiency2004}.

\subsection{Phase aberrations}
\begin{figure}
\begin{center}
\includegraphics[width=83mm]{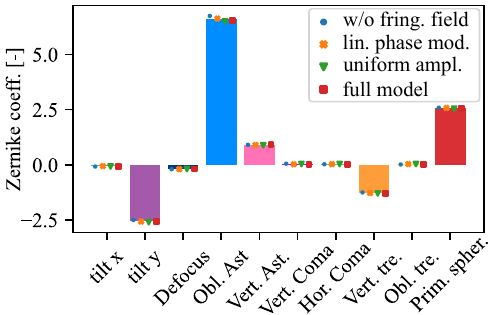}
\end{center}
\caption 
{ \label{fig:zernike}
Derived phase aberrations represented by Zernike coefficients. The bar plots show the mean value of the four calibrations while the markers show the values of the different calibrations.} 
\end{figure}

The derived Zernike coefficients are given in \autoref{fig:zernike}. From the results in \autoref{tab:Table1} and \autoref{fig:results} we can see that phase aberrations have the strongest influence on image quality. The results from \autoref{fig:zernike} show that the highest aberrations in the system are the defocus coefficient and the primary spherical aberration. The derived Zernike coefficients show high consistency over the four calibrations. The phase distributions for modeling the higher spatial frequencies show values between $-\pi/2$ and $+\pi/2$ and low deviations between the different calibrations.

\subsection{Illumination amplitude}
\begin{figure}
\begin{center}
\includegraphics[width=83mm]{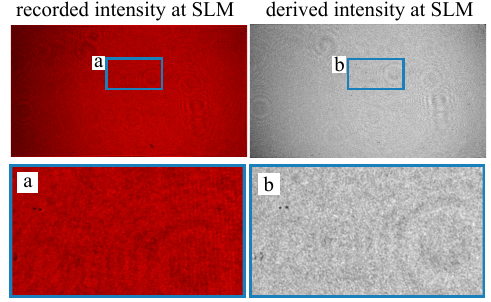}
\end{center}
\caption 
{ \label{fig:ampl}
Comparison of the learned intensity at the SLM with the recorded intensity at the SLM. Dust particles on the SLM and diffraction patterns from particles on lenses match between recording and derived intensity. The dark spots stem from the same dust particle on the cover glass of the SLM due to the tilt of the SLM.} 
\end{figure}

In our case, the illumination amplitude has the lowest effect on image quality, and using a uniform illumination amplitude even shows a better SSIM value than the full model. Even tough the learned amplitude resembles the real amplitude very well, as dust particles on the SLM are correctly located in the learned amplitude as seen in \autoref{fig:ampl}, it shows some degree of noise for uniformly illuminated areas. 
This noise can lead to undesired diffracted light when phase patterns are computed. The effect of the undesired light diffraction is only marginal, but has a strong effect on the SSIM metric, when the area should be completely black but contains low levels of noise.

\section{Conclusion}
We demonstrated the influence of different aberrations in combination with a CITL calibration for far-field holograms. When certain aberrations are excluded, their effects can be partially compensated by others. For example, if the impact of non-linear phase modulation is omitted, the maximum phase delay is instead learned from the fringing-field kernel. The highest PSNR and lowest speckle contrast are achieved with the calibration model that includes all aberrations. Among the different types, phase aberrations proved to be the most prominent in our system.

We further point out that the results suggest CITL calibration could provide a solution for phase retrieval methods, provided the origins of the aberrations can be separated, for example, into those introduced by specific lenses and by the SLM.

\begin{acknowledgement}

\end{acknowledgement}

\begin{funding}

\end{funding}
\bibliographystyle{IEEEtran}
\bibliography{TM_Paper}

\section*{Author information}

\vspace{2ex} 
\begin{figure}[!ht]
    \includegraphics[width=0.3\linewidth]{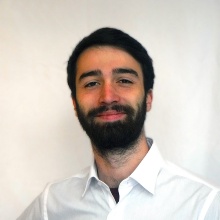}
\end{figure}
\noindent
Institut für Technische Optik, Universität
Stuttgart, Pfaffenwaldring 9, 70569 \\
Stuttgart, Germany\\
zimmermann@ito.uni-stuttgart.de\\
Markus Zimmermann is a PhD student at the Institut für Technische Optik at the University of Stuttgart. Here he completed his master's degree in photonic engineering in 2020. His area of expertise is in holographic display technology.

\vspace{2ex}
\begin{figure}[!ht]
    \includegraphics[width=0.3\linewidth]{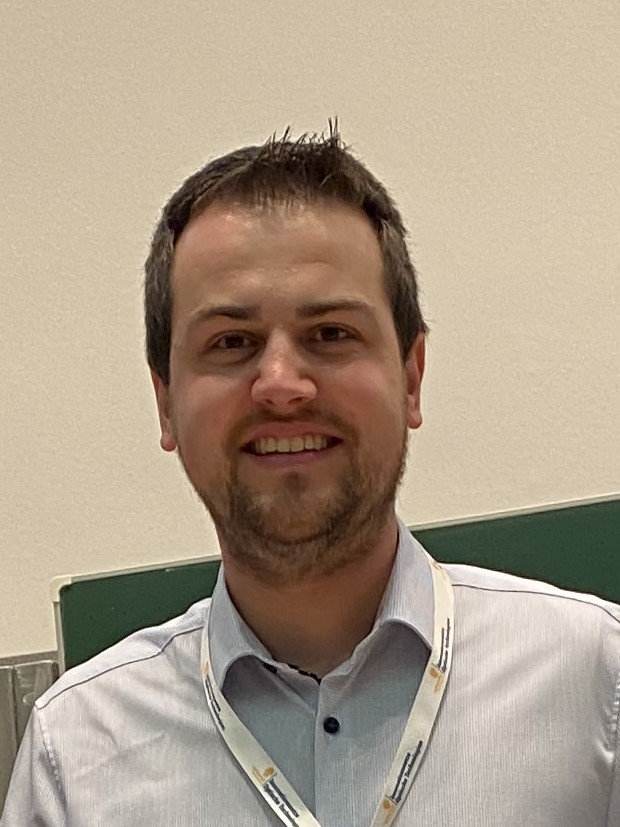}
\end{figure}
\noindent Institut für Technische Optik, Universität
Stuttgart, Pfaffenwaldring 9, 70569 \\
Stuttgart, Germany\\
brenner@ito.uni-stuttgart.de\\
Andreas Brenner is a PhD student at the Institut für Technische Optik at the University of Stuttgart. He received his master’s degree in 2024 from the University of Stuttgart in photonic engineering. His main research interests are in the field of holographic display technologies and interferometry.

\vspace{2ex}
\begin{figure}[!ht]
    \includegraphics[width=0.3\linewidth]{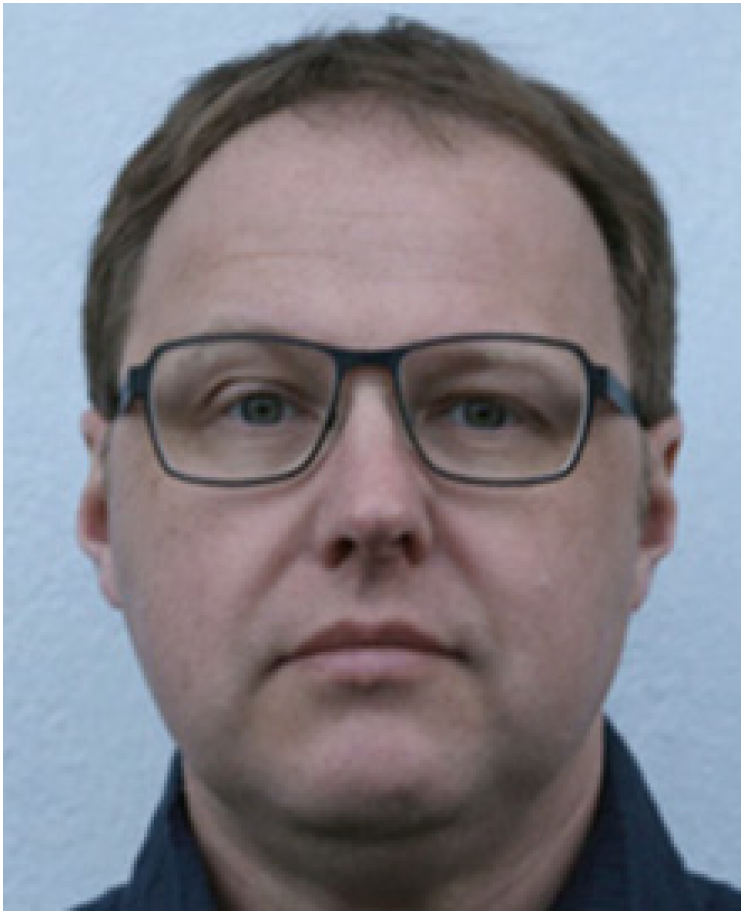}
\end{figure}
\noindent Institut für Technische Optik, Universität
Stuttgart, Pfaffenwaldring 9, 70569 \\
Stuttgart, Germany\\
haist@ito.uni-stuttgart.de\\
Tobias Haist studied physics and received his PhD in engineering from the University of Stuttgart. Currently, he is leading the group 3D Surface Metrology at the Institut für Technische Optik, where he is working on new applications for spatial light modulators and 3-D measurement systems. His main research interests include optical and digital image processing, computer-generated holography, and optical measurement systems.

\vspace{2ex}
\begin{figure}[!ht]
    \includegraphics[width=0.3\linewidth]{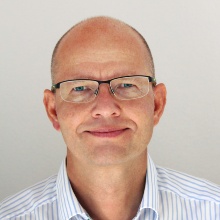}
\end{figure}
\noindent Institut für Technische Optik, Universität
Stuttgart, Pfaffenwaldring 9, 70569 \\
Stuttgart, Germany\\
reichelt@ito.uni-stuttgart.de\\ 
Stephan Reichelt is a professor at the University of Stuttgart and head of the Institute of Applied Optics. He received his PhD from the University of Stuttgart in 2004 in the field of interferometric testing of aspherical optics and computer-generated holograms. He has authored or co-authored more than 60 technical publications and holds several patents. His current research focuses on optical metrology, inspection, and sensing for industrial and biomedical applications. He is a member of Optica, SPIE, the European Optical Society, and the German Society for Applied Optics.

\end{document}